\documentclass{article}

\begin{document}

\title{Quantum particle production\\
near the big rip revisited}
\author{Fl\'avio G. Alvarenga\\
Departamento de Engenharia e Ci\^encias Exatas - CEUNES\\
Universidade Federal do Esp\'{\i}rito Santo\\
CEP 29933-415 S\~ao Mateus, Esp\'{\i}rito Santo, Brasil\\
\\
Antonio B. Batista, J\'{u}lio C.
Fabris and St\'{e}phane Houndjo \\
Departamento de F\'{\i}sica, \\
Universidade Federal do Esp\'{\i}rito Santo, \\
CEP 29060-900 Vit\'{o}ria, Esp\'{\i}rito Santo, Brasil}
\date{}
\maketitle

\begin{abstract}
The effect of quantum particle production near the big rip singularity
has been investigated previously, with the conclusion that the energy of
the produced particle decreases as the future singularity is approached.
Hence, the effect of particle production would not be effective to avoid
the big rip singularity. That calculation was performed by introducing an ultra-violet
cut-off. In the present work we consider a renormalization of the energy-momentum
tensor, obtainning a different expression for the particle production. The new expression
seems to indicate that the effect of particle production may be dominant as the
singularity is approached.

\end{abstract}

\vspace{0.5cm} \leftline{PACS: 98.80.-k, 04.62.+v}

\section{Introduction}

The particle creation in curved space-time is a quantum effect associated with the non-unicity of the vacuum state when the space-time does not have a time-like killing vector
\cite{birrel,grib}.
In this situation the curvature changes with time, and the characterization of the vacuum
state depends on time. Hence, if we fix an initial vacuum state at, let us say, $t = 0$, this
vaccum state becomes later a multiparticle state. This is just another way of saying the
particles are created as the space-time evolves with time.
\par
There are two main situations where the particle creation
as a gravitational phenomena has been especially studied: the space-time of a black hole and
the expanding universe. If we consider the fact that the black hole must be formed dynamically by the gravitational collapse process, there is no time invariance of the
configuration, the curvature grows with time, and particle must be created as the event
horizon is formed. On the other hand, the time invariance does not exist when
the universe expands, and we must also expect that the phenomena of particle creation must
appear in the
usual cosmological scenarios. We must expect that the particle creation rate must
be linked with the curvature of the four-dimensional space-time.
\par
Considering the usual Friedmann universe, with homogeneity and isotropy, there are
two situations where the particle creation phenomena does not occur. The first one, is
when the universe is in a de Sitter state. The de Sitter space-time is maximally symmetric.
Hence, it contains a timelike Killing vector, and the vacuum state must be unique. The other
situation is a radiative universe. Since the electromagnetic field is conformal invariant,
the particle production is zero; in fact, an isotropic conformal field does not lead
to any particle creation.
\par
For other types of Friedmann universes, the situation is more complex. In the cases where
the strong energy condition $\rho + 3p \geq 0$ ($\rho$ the energy density and
$p$ the pressure) is satisfied, the curvature is initially infinite, decreasing
later with time. In such a situation, the greastet problem is to define an initial vacuum
state: such definition becomes quite arbitrary, and the vacuum state appears naturally only
in the asymptotic future. When the strong energy condition is violated, however, it is
possible to define uniquely a vacuum state, and the problem of particle creation seems
to be well posed.
\par
The problem of particle creation in a Friedmann universe has been analysed recently in
reference \cite{brasil1}. The main motivation of that analysis is the fact that the universe
today seems to be dominated by a phantom fluid \cite{spergel}, a fluid
that violates the dominant energy condition $\rho + p \geq 0$. A universe
dominated by a phantom fluid will end inevitably in a future singurality, in a finite
proper time, called big-rip: as the universe expands, the energy density of the
phantom fluid increases, becoming divergent at a given moment in the future \cite{caldwell}. If this is so, the question it was
tried to be answered in that work was: can the particle creation to be
so effective that it leads to the avoidance of
the big rip? It was used a massless scalar field leaving in a universe dominated by
a fluid with an equation of state $p = \alpha\rho$. In particular, it was payed special
attention to the case $\alpha < - 1/3$, the situation where the strong energy condition
is violated.
\par
The analysis of this problem becomes cumbersome due to the fact that the energy-momentum
tensor for the scalar field becomes divergent in the ultraviolet limit. Hence, a renormalization procedure must be employed. In the case of curved space-time this is
not a simple task. In reference \cite{brasil1} it was chosen to introduce an ultraviolet cut-off connected with the Planckian frequency. The reason for that is the fact that
such problem involves gravitation, and above the Planck frequency we enter in the quantum
gravity domain, and very probable the dispersion relation connecting the frequency $\omega$ and
the wavenumber $k$ must be changed.
\par
The answer for the question of the avoidance of the big-rip due to particle creation found
in \cite{brasil1} using the procedure described above was negative: the number
of particle created goes to infinity, but the associated energy density goes to zero, as
the big-rip is approached. This seems to agree with the result obtaining in reference
\cite{brasil2}, where a similar study was performed in the case of the sudden singularity
\cite{barrow},
a singularity that occurs also in the future but without violation neither of the strong
energy condition nor the dominant energy condition: the sudden singularity is robust with
respect to the particle creation phenomena.
\par
However, it must be remarked that, in contrast with what happens in the case of reference
\cite{brasil1}, in the sudden singularity scenario the subtraction of the infinities in
the energy-momentum tensor is a much easier task, and is possible to obtain a renormalized
energy-momentum tensor (ignoring all transplanckian problems). Here, we would like to come
back to the problem treated in reference \cite{brasil1}, and ignore the transplanckian problem also in big-rip scenario, trying to renormalize the energy-momentum tensor for
the massless scalar field. We employ the $n$-wave method, which is quite convinient
for an isotropic universe. However, this task becomes more complicated due to the existence of logarithmic divergence in the ultraviolet limit. We propose a simple way to deal with this problem, without introducing any
new arbitray parameter.
We obtain a finite energy-momentum tensor. However, using now this renormalized energy, we
verify that the energy associated with the particle production diverges near the big
rip, and it can become the dominant energy component, altering the final state of the universe. The unicity of the answer and its modification due to transplanckian considerations
remain open questions. However, our result points out that the problem can be much more
complex than it seems at first glance.
\par
This paper is organized as follows. In next section we revise the general problem
of renormalization of a quantum scalar field living in a FRW space-time. In section
3 we apply the formalism to the case of a radiative universe, and in section
4 to the case of a de Sitter universe. In section 5 the general inflationary case
is discussed, showing that the method has some limitations due to the appearence
of an ultraviolet divergence. A full renormalisation is proposed in section 6, obtainning
as result that the big rip may be avoid due to quantum effects. In section 7 we present our
conclusions.

\section{General problem of renormalization}

The Friedmann-Lema\^{\i}tre-Robertson-Walker (FLRW) metric
\begin{equation}
 ds^2 = dt^2 - a(t)^2(dx^2 + dy^2 + dz^2) 
\end{equation}
describes a spatially flat, expanding universe, with $a(t)$ being the time-dependent
scale factor. In terms of the conformal time defined by $dt = ad\eta$, the FLRW metric
takes the form
\begin{equation}
ds^2 = a^2(d\eta^2 - dx^2 - dy^2 - dz^2) .
\end{equation}
If the universe is filled with a barotropic perfect fluid, with the equation of state
$p = \alpha\rho$, the Friedmand equation
\begin{equation}
 \biggr(\frac{a'}{a}\biggl)^2 = \frac{8\pi G}{3}\rho\,a^2
\end{equation}
(primes meaning derivative with respect to the conformal time $\eta$)
implies
\begin{equation}
a \propto \eta^\beta , \quad \beta = \frac{2}{1 + 3\alpha} .
\end{equation}
For $\alpha > - 1/3$, the strong energy condition $\rho + 3p \geq 0$ is satisfied and
the universe expands desacelerating as $\eta \rightarrow \infty$; for $\alpha < - 1/3$ the
universe expands accelerating as $\eta \rightarrow 0_-$. In particular, $\alpha = - 1$ corresponds to a de Sitter universe, while $\alpha = 1/3$ implies a radiation-dominated
universe. If $\alpha < - 1$, the fluid is called phantom, and the universe must
faces a big-rip in the future.
\par
The fundamental equation for a massless scalar field is living in a FLRW space-time is,
\begin{equation}
\label{fun}
{\phi}''_k + 2\frac{a'}{a}\phi'_k + k^2\phi_k = 0  .
\end{equation}
The general expressions for the energy and pressure of the quantum scalar field are given by the following expressions:
\begin{eqnarray}
\rho = \frac{1}{a^2}\int_0^\infty\biggr\{{{\phi}'}_k{{\phi}'}_k^{*} + k^2\phi_k\phi_k^{*}\biggl\}k^2dk  ; \\
p =\frac{1}{a^2} \int_0^\infty\biggr\{{{\phi}'}_k{{\phi}'}_k^{*} - \frac{k^2}{3}\phi_k\phi_{k^{*}}\biggl\}k^2dk  .
\end{eqnarray}
These expressions can be written as
\begin{eqnarray}
\label{e1}
\rho &=& \int_0^\infty k^2 \,E_k\,dk , \quad E_k = \frac{1}{a^2}\biggr\{{{\phi}'}_k{{\phi}'}_k^{*} + k^2\phi_k\phi_k^{*}\biggl\} ,\\
\label{e2}
p &=& \int_0^\infty k^2\,P_k\,dk , \quad P_k = \frac{1}{a^2}\biggr\{{{\phi}'}_k{{\phi}'}_k^{*} - \frac{k^2}{3}\phi_k\phi_{k^{*}}\biggl\}  .
\end{eqnarray}
\par
In general, the above expressions for the energy and the pressure have quartic, quadratic and logarithmic divergencies in the ultraviolet limit, while it is regular in the infrared limit.
In order to cope with these divergencies, we employ the $n$-wave method
described in the reference \cite{zel}. Essentially, this method consists in subtracting
terms obtainned by expanding $E_k$ and $P_k$ in powers of $k^{-2}$: 
\begin{eqnarray}
\label{en1}
E_k^{ren} = E_k - E_k^0 - E_k^1 - \frac{1}{2}E_k^2  , \\
\label{en2}
P_k^{ren} = P_k - P_k^0 - P_k^1 - \frac{1}{2}P_k^2  , 
\end{eqnarray}
where 
\begin{equation}
E_k^p = \lim_{n\rightarrow \infty} \frac{\partial^p E^n_k}{\partial{(n^{-2}})^p} , \quad P_k^p = \lim_{n\rightarrow \infty} \frac{\partial^p P^n_k}{\partial{(n^{-2}})^p}  ,
\end{equation}
with the definitions,
\begin{equation}
\label{equa1}
E^n_k = \frac{1}{n}E_k(nk) , \quad P^n_k = \frac{1}{n}P_k(nk) .
\end{equation}
It will come out more convient to express the derivatives as
\begin{equation}
\frac{\partial f}{\partial n^{-2}} = \frac{\partial n}{\partial n^{-2}}\frac{\partial f}{\partial n} = - \frac{n^3}{2}\frac{\partial f}{\partial n} ,
\end{equation}
and subsequently for the higher derivatives, $f$ being either $E_k$ or $P_k$. Hence, we find:
\begin{eqnarray}
\frac{\partial f}{\partial n^{-2}} &=& - \frac{n^3}{2}\frac{\partial f}{\partial n} , \\
\frac{\partial^2 f}{\partial {n^{-2}}^2} &=& \frac{3}{4}n^5\frac{\partial f}{\partial n} + \frac{n^6}{4}\frac{\partial^2 f}{\partial n^2} .
\end{eqnarray}
\par
It will be necessary later to use the following expressions for the Hankel's functions in the limit of large values for the argument:
\begin{equation}
\label{expan1}
H^{(1)}_\nu(x) \sim \sqrt{\frac{2}{\pi x}}\biggr\{P(\nu,x) + iQ(\nu,x)\biggl\}e^{i\chi}  ;
\end{equation}
\begin{equation}
\label{expan2}
H^{(2)}_\nu(x) \sim \sqrt{\frac{2}{\pi x}}\biggr\{P(\nu,x) - iQ(\nu,x)\biggl\}e^{-i\chi}  , \\
\end{equation}
where $\chi = \bigg[x-\bigg(\nu+\frac{1}{2}\bigg)\frac{\pi}{2}\bigg]$ and
\begin{eqnarray}
P(\nu,x) &=& \sum_{k=0}^\infty (-1)^k\frac{(\nu,2k)}{(2x)^{2k}} = 1 - \frac{(4\nu^2 - 1)(4\nu^2 - 9)}{2!(8x)^2} \nonumber\\
&+&  \frac{(4\nu^2 - 1)(4\nu^2 - 9)(4\nu^2 - 25)(4\nu^2 - 49)}{4!(8x)^4} + ...  ,\\
Q(\nu,x) &=& \sum_{k=0}^\infty (-1)^k\frac{(\nu,2k+1)}{(2x)^{2k+1}} = \frac{4\nu^2 - 1}{8x } \nonumber\\
&-& \frac{(4\nu^2 - 1)(4\nu^2 - 9)(4\nu^2 - 25)}{3!(8x)^3} + ...  ,\\
(\nu,k) &=& \frac{\Gamma(1/2 + n + k)}{k!\Gamma(1/2 + n - k)}  .
\end{eqnarray}
\par
We will consider from now
on two particular cases, the radiative case and the de Sitter case, as
well as the general inflationary and phantom cases.

\section{Radiative case}

Let us consider now the particular case of the the flat universe dominated by the
radiative fluid. Since the
radiative fluid is conformal invariant, and the universe is isotropic and homogenous,
we must expect that the rate of particle creation is zero, as it is stated in reference
\cite{zel} (see also reference \cite{parker}).

For a radiative universe, the
scale factor behaves as $a = a_0\eta$. Hence, the equation (\ref{fun}) reads now,
\begin{equation}
\label{er}
\phi'' + 2\frac{\phi'}{\eta} + k^2\phi = 0  .
\end{equation}
The general solution is
\begin{equation}
\phi = c_1(k)\frac{e^{i(k\eta - \vec k\cdot\vec x)}}{\eta} + c_2(k)\frac{e^{i(k\eta + \vec k\cdot\vec x)}}{\eta}  .
\end{equation}
We find the initial Bunch-Davies vacuum if, for example,
\begin{equation}
c_1 = \sqrt{\frac{1}{2k}} , \quad c_2 = 0  .
\end{equation}
Now the energy density and the pressure read,
\begin{eqnarray}
\rho &=& \frac{1}{2a^2}\int_0^\infty\biggr\{2k^2 + \frac{1}{\eta^2}\biggl\}\frac{k}{\eta^2}dk = \frac{1}{2a_0^2}\int_0^\infty\biggr\{2k^2 + \frac{1}{\eta^2}\biggl\}\frac{k}{\eta^4}dk   ,\\
p &=& \frac{1}{2a^2}\int_0^\infty\biggr\{\frac{2}{3}k^2 + \frac{1}{\eta^2}\biggl\}\frac{k}{\eta^2}dk =  \frac{1}{2a_0^2}\int_0^\infty\biggr\{\frac{2}{3}k^2 + \frac{1}{\eta^2}\biggl\}\frac{k}{\eta^4}dk  .
\end{eqnarray}
These expressions for the energy and pressure obey the conservation law,
\begin{equation}
\rho' + 3\frac{a'}{a}(\rho + p) = 0  .
\end{equation}
\par
From (\ref{e1},\ref{e2}), we find:
\begin{eqnarray}
E_k &=& \biggr\{2k^2 + \frac{1}{\eta^2}\biggl\}\frac{1}{k\eta^4}  , \\
P_k &=& \biggr\{\frac{2}{3}k^2 + \frac{1}{\eta^2}\biggl\}\frac{1}{k\eta^4}  .
\end{eqnarray}
There is a quartic and a quadratic divergences both for the energy and for the pressure.
\par
In order to give sense to the energy and pressure expression, let us proceed with the
renormalization schema described in section $2$.
Using the expressions for $E_k^p$ e $P_k^p$, we obtain the following relations:
\begin{eqnarray}
E_k^0 = 2\frac{k}{\eta^4} , \quad E_k^1 = \frac{1}{k\eta^6} , \quad E_k^2 = 0 , \\
P_k^0 = \frac{2}{3}\frac{k}{\eta^4} , \quad P_k^1 = \frac{1}{k\eta^6} , \quad P_k^2 = 0   
\end{eqnarray}
Hence, we have,
\begin{eqnarray}
E_k^{ren} &=& E_k - E_k^0 - E_k^1 - \frac{1}{2}E_k^2 = 0 ,  \\
P_k^{ren} &=& P_k - P_k^0 - P_k^1 - \frac{1}{2}P_k^2 = 0  .
\end{eqnarray}
\par
Consequently, the initial expression for the energy diverges but, after renormalization, it is equal to zero, in agreement with the fact that no particle is created during the
radiative phase.
\par
Note that the equation (\ref{er}), after the redefinition $\phi = \frac{\xi}{a}$, takes the
form
\begin{equation}
 \xi'' + k^2\xi = 0 .
\end{equation}
From this expression it is clear that we could expect no particle creation from the begining.

\section{Inflation: de Sitter phase}

In a de Sitter phase, resulting from the equation of state $p = - \rho$ ($\alpha = - 1$), the scale factor is given by
\begin{eqnarray}
a = - \frac{1}{\eta}  .
\end{eqnarray}
The Klein-Gordon equation reads,
\begin{eqnarray}
\phi'' - 2\frac{\phi'}{\eta} + k^2\phi = 0  .
\end{eqnarray}
The solution is,
\begin{eqnarray}
\phi = \eta^{3/2}\biggr\{c_1\,H_{3/2}^{(1)}(k\eta) + c_2\,H_{3/2}^{(2)}(k\eta)\biggl\}  .
\end{eqnarray}
Notice that, from now on, we have made the substitution $\eta \rightarrow - \eta$, in order
to avoid the repetivie use of the absolute value of the original $\eta$ parameter which is defined
in the negative real axis.
\par
The initial condition is imposed at $\eta \rightarrow \infty$, where
\begin{equation}
\phi = \eta^{3/2}\sqrt{\frac{2}{\pi k\eta}}\biggr\{c_1\,e^{i(k\eta - \pi)} + c_2\,e^{-i(k\eta - \pi)}\biggl\}  .
\end{equation}
We can obtain an initial Bunch-Davies vacuum state if
\begin{equation}
c_1 = - \frac{\sqrt{\pi}}{2} , \quad c_2 = 0  .
\end{equation}
Hence, we have the final solution
\begin{equation}
\phi = \eta^{3/2}c_1\,H_{3/2}^{(1)}(k\eta)  .
\end{equation}
\par
Using the recurrence relations for the Hankel's functions, we obtain the following expression for the energy and the pressure:
\begin{eqnarray}
\rho &=& 4\pi c_1^2\eta^{5}\int_0^\infty k^4\biggr\{H_{1/2}^{(1)}(k\eta)\,H_{1/2}^{(2)}(k\eta) + H_{3/2}^{(1)}(k\eta)\,H_{3/2}^{(2)}(k\eta)\biggl\}dk  , \\
p &=& 4\pi c_1^2\eta^{5}\int_0^\infty k^4\biggr\{H_{1/2}^{(1)}(k\eta)\,H_{1/2}^{(2)}(k\eta) - \frac{1}{3} H_{3/2}^{(1)}(k\eta)\,H_{3/2}^{(2)}(k\eta)\biggl\}dk
 .
\end{eqnarray}
These expressions contain a quartic and a quadratic divergences as in the radiative case.
In fact, these expressions can be simplified using the following forms for the Hankel's functions:
\begin{eqnarray}
H^{(1)}_{1/2}(x) = - i\sqrt{\frac{2}{\pi x}}e^{ix} &,& \quad H^{(2)}_{1/2}(x) = i\sqrt{\frac{2}{\pi x}}e^{-ix}  ;\\
H^{(1)}_{3/2}(x) = - \sqrt{\frac{2}{\pi x}}\biggr\{1 + \frac{i}{x}\biggl\}e^{ix} &,& \quad H^{(2)}_{3/2}(x) = - \sqrt{\frac{2}{\pi x}}\biggr\{1 - \frac{i}{x}\biggl\}e^{-ix}  .
\end{eqnarray}
With these expressions for the Hankel's functions, the expression for the energy and the pressure reduces to:
\begin{eqnarray}
\rho &=& 8c_1^2\eta^4\int_0^\infty k^3\biggr\{2 + \frac{1}{(k\eta)^2}\biggl\}dk  ;\\
p &=& \frac{8}{3}c_1^2\eta^4\int_0^\infty k^3\biggr\{2 - \frac{1}{(k\eta)^2}\biggl\}dk  .
\end{eqnarray}
We see to appear the usual quartic and quadratic divergencies in the ultraviolet limit. There is no divergence in infrared limit.
\par
We proceed with the renormalization of the energy and the pressure associated with the
quantum field.
\begin{itemize}
\item Renormalization of the energy.
\begin{eqnarray}
E_k^n &=& \frac{\eta^4}{2\pi^2}\biggr\{2k + \frac{1}{k\eta^2n^2}\biggl\} ; \\
\frac{\partial E_k^n}{\partial n^{-2}} &=& \frac{\eta^2}{2\pi^2 k}  ;\\
\frac{\partial^2 E^n_k}{\partial {n^{-2}}^2} &=& 0   .
\end{eqnarray}
In the limit $n \rightarrow \infty$, we have
\begin{eqnarray}
E_k^0 = \lim_{n\rightarrow\infty}E_k^n &=& \frac{\eta^4k}{\pi^2} ; \\
E_k^1 = \lim_{n\rightarrow\infty}\frac{\partial E_k^n}{\partial n^{-2}} &=& \frac{\eta^2}{2\pi^2 k} ;\\
E_k^2 = \lim_{n\rightarrow\infty}\frac{\partial^2 \rho^n_k}{\partial {n^{-2}}^2} &=& 0  .
\end{eqnarray}
Hence, the renormalized energy is:
\begin{equation}
E_k^{ren} = \frac{2}{\pi}\eta^4\biggr\{2k + \frac{1}{k\eta^2} - 2k - \frac{1}{k\eta^2}\biggl\} \equiv 0 .
\end{equation}
\item Renormalization of the pressure:
\begin{eqnarray}
P_k^n &=& \frac{\eta^4}{6\pi^2}\biggr\{2k - \frac{1}{k\eta^2 n^2}\biggl\} ; \\
\frac{\partial P_k^n}{\partial n^{-2}} &=& - \frac{\eta^2}{6\pi^2 k}  ;\\
\frac{\partial^2 P^n_k}{\partial {n^{-2}}^2} &=& 0  .
\end{eqnarray}
In the limit $n \rightarrow \infty$, we have
\begin{eqnarray}
P_k^0 = \lim_{n\rightarrow\infty}P_k^n &=& \frac{\eta^4k}{3\pi^2} ; \\
P_k^1 = \lim_{n\rightarrow\infty}\frac{\partial P_k^n}{\partial n^{-2}} &=& - \frac{\eta^2}{6\pi^2 k}  ;\\
P_k^2 = \lim_{n\rightarrow\infty}\frac{\partial^2 P^n_k}{\partial {n^{-2}}^2} &=& 0   .
\end{eqnarray}
Hence, the renormalized pressure is:
\begin{equation}
P_k^{ren} = \frac{2}{3\pi}\eta^4\biggr\{2k - \frac{1}{k\eta^2} - 2k + \frac{1}{k\eta^2}\biggl\} \equiv 0  .
\end{equation}
\end{itemize}
\par
Again this is a result that we could expect from the begining since the de Sitter space-time
has maximal symmetry. As consequence, there is a time-like Killing vector, and no
particle can be created, and the energy and pressure associated to the quantum field must
be zero.

\section{Inflation: the general case}

Let us consider now the general inflationary case.
The scale factor reads,
\begin{equation}
a = a_0\eta^{2/(1 + 3\alpha)}  .
\end{equation}
There is inflation if $\alpha < - 1/3$. 
The Klein-Gordon equation reads,
\begin{equation}
\phi'' + 2\beta\frac{\phi'}{\eta} + k^2\phi = 0 , \quad \beta = \frac{2}{1 + 3\alpha}  .
\end{equation}
The general solution is
\begin{equation}
\phi = \eta^\nu\biggr\{c_1H_\nu^{(1)}(k\eta) + c_2H_\nu^{(2)}(k\eta)\biggl\} , \quad \nu = \frac{1}{2} - \beta .
\end{equation}
As before, we can choose $c_1$ in order to have the Bunch-Davies vacuum in the limit $k \rightarrow \infty$, and $c_2 = 0$.
Hence, the final expression is
\begin{equation}
\phi = \eta^\nu c_1H_\nu^{(1)}(k\eta)  ,
\end{equation}
where $c_1$ does not depend on $k$.
\par
The corresponding expressions for the energy and pressure are
\begin{eqnarray}
\label{energia}
\rho &=& A\eta^{2\nu - 2\beta}\int_0^\infty k^4\biggr\{H_{\nu - 1}^{(1)}(k\eta)\,H_{\nu - 1}^{(2)}(k\eta) + H_\nu^{(1)}(k\eta)\,H_\nu^{(2)}(k\eta)\biggl\}dk , \\
\label{pressao}
p &=& A \eta^{2\nu - 2\beta}\int_0^\infty k^4\biggr\{H_{\nu - 1}^{(1)}(k\eta)\,H_{\nu - 1}^{(2)}(k\eta) -  \frac{1}{3}H_\nu^{(1)}(k\eta)\,H_\nu^{(2)}(k\eta)\biggl\}dk ,
\end{eqnarray}
where $A=4\pi c_1^2$.
\par
The conservation law 
\begin{equation}
{\rho}'+ 3\frac{\beta}{\eta}(\rho + p) = 0 \, 
\end{equation}
is verified. 
\par
As in the preceding cases, there is no divergence in the infrared limit.
Let us find first the divergences that may occur at the limit $k \rightarrow \infty$.
Using the expansion (\ref{expan1},\ref{expan2}), we must go until order $k^{-5}$ in the brackets of the integrands of (\ref{energia},\ref{pressao}): due to the term $k^4$ in the
integral of the energy and pressure, all terms up to this
order in the product of Hankel's functions are divergents. In particular, the term of order $k^{-5}$ gives a logarithmic divergence, the term $k^{-3}$ a quadratic divergence, and the term $k^{-1}$ a quartic divergence, and
all subsequent terms in the expansion are zero in the limit $k \rightarrow \infty$. Note that in both examples treated before, the logarithmic divergence is absent.
\par
Using (\ref{expan1},\ref{expan2}), we find
\begin{eqnarray}
\label{a3}
H_{\nu}^{(1)}(x)H_{\nu-1 }^{(2)}(x) &=& \frac{2}{\pi x}[-iP_{\nu}(x)P_{\nu-1}(x) - P_{\nu}(x)Q_{\nu-1}(x) + \nonumber\\
Q_{\nu}(x) P_{\nu -1}(x)
&-& iQ_{\nu}(x)Q_{\nu -1}(x)] ; \\
H_{\nu -1}^{(1)}(x)H_{\nu}^{(2)}(x) &=& \frac{2}{\pi x}[iP_{\nu-1}(x)P_{\nu}(x) + P_{\nu-1}(x)Q_{\nu}(x) - \nonumber \\
P_{\nu}(x) Q_{\nu -1}(x)
\label{a4}
&+& iQ_{\nu}(x)Q_{\nu -1}(x)] ;\\
H_{\nu - 1}^{(1)}(x)H_{\nu }^{(2)}(x) + H_{\nu}^{(1)}(x)H_{\nu-1}^{(2)}(x) &=& \frac{4}{\pi x}[Q_{\nu}(x)P_{\nu-1}(x) -
\nonumber\\ && P_{\nu}(x)Q_{\nu-1}(x)]  .
\end{eqnarray}

Using the expansion for the functions $P$ and $Q$, the expression for the energy and the pressure become:
\begin{eqnarray}
E_k &=& A\eta^{2\nu - 2\beta - 2}\frac{2}{\pi}x\biggr\{2 + \frac{(2\nu - 1)^2}{4\,x^2} \nonumber\\
&+& 3\frac{(2\nu - 3)(2\nu - 1)^2(2\nu +1)}{64\,x^4} + \cdot\cdot\cdot\biggl\} , \\
P_k &=& A\eta^{2\nu - 2\beta - 2}\frac{2}{\pi}x\biggr\{\frac{2}{3} + \frac{4\nu^2 - 12\nu + 5}{12\,x^2} \nonumber\\
&+& \frac{(2\nu - 3)(2\nu - 1)(2\nu - 9)(2\nu +1)}{64\,x^4} + \cdot\cdot\cdot\biggl\} ,
\end{eqnarray}
where $x = k\eta$.
It is clear, at this approximation, that there are a quadratic, a quartic and a logarithmic
divergences in the ultraviolet limit. Let us compute the different components to be
subtracted from the above expressions, as we have done before.

\subsection{Computation of the energy}

The energy reads,
\begin{equation}
E_{k} = \eta^{\gamma}k^2\biggr\{H_{\nu - 1}^{(1)}(k\eta)\,H_{\nu - 1}^{(2)}(k\eta) + H_\nu^{(1)}(k\eta)\,H_\nu^{(2)}(k\eta)\biggl\}\, ,
\end{equation}
where $\gamma=2(\nu-\beta)$.
As consequence, we find
\begin{equation}
E_{k}^{n} = \eta^{\gamma}n k^2\biggr\{H_{\nu - 1}^{(1)}(nk\eta)\,H_{\nu - 1}^{(2)}(nk\eta) + H_\nu^{(1)}(nk\eta)\,H_\nu^{(2)}(nk\eta)\biggl\}\, ,
\end{equation}
where
$x \rightarrow x=nk\eta$.
\par
Consequently, the expression for the energy can be rewritten as:
\begin{equation}
E_k^n = \eta^{\gamma - 1}k\,f(x) , \quad f(x) = x\biggr\{H_{\nu - 1}^{(1)}(x)H_{\nu - 1}^{(2)}(x) + H_\nu^{(1)}(x)H_\nu^{(2)}(x)\biggl\} . 
\end{equation}
Hence, we find:
\begin{eqnarray}
\label{stephan1}
E^0_k &=& \lim_{x\rightarrow\infty}\eta^{\gamma - 1} kx\biggr\{H_{\nu - 1}^{(1)}(x)H_{\nu - 1}^{(2)}(x) + H_\nu^{(1)}(x)H_\nu^{(2)}(x)\biggl\}  ; \\
\label{stephan2}
E^1_k &=& \lim_{x\rightarrow\infty}\biggr\{- \frac{2\nu - 1}{2k}\eta^{\gamma - 3}x^3\biggr[H_{\nu - 1}^{(1)}(x)H_{\nu - 1}^{(2)} - H_\nu^{(1)}(x)H_\nu^{(2)}(x)\biggl]\biggl\} ;\\
\label{flavio1}
E^2_k &=& \lim_{x\rightarrow\infty}\frac{2\nu - 1}{4k^3}\eta^{\gamma - 5}\,x^5\biggr\{(2\nu +1 )H^{(1)}_{\nu-1}(x)H^{(2)}_{\nu - 1}(x) +
(2\nu - 3){H_{\nu}}^{(1)}(x){H_{\nu}}^{(2)}(x)\nonumber\\
&-& 2x\biggr[H_\nu^{(1)}(x)H_{\nu-1}^{(2)}(x) + H_{\nu - 1}^{(1)}(x)H_\nu^{(2)}(x)\biggl]\biggl\} .
\end{eqnarray}
\par
Now we are read to compute $E^{0}_{k}$, $E^{1}_{k}$, $E^{2}_{k}$ and  $E_{k}$. Using first the relation
\begin{eqnarray}
& &H^{(1)}_{\nu-1}(x)H^{(2)}_{\nu-1}(x)+ H^{(1)}_{\nu}(x)H^{(2)}_{\nu}(x) =
\nonumber \\
&&\frac{2}{\pi x}\biggr\{2+\frac{4\nu^4-4\nu+1}{4x^2}+\frac{48\nu^4-96\nu^3+24\nu^2+24\nu-9}{64 x^4}\biggl\}\, , \\
&&H^{(1)}_\nu(x)H^{(2)}_{\nu - 1}(x) + H^{(1)}_{\nu - 1}(x)H^{(2)}_\nu(x) = \nonumber \\
&&\frac{2}{\pi x}\biggr\{\frac{2\nu - 1}{x} + 4\frac{384\nu^3 - 576\nu^2 - 96\nu + 144}{3(8x)^3} -\nonumber\\
&&\frac{1}{3(8x)^5}\biggr[(4\nu^2 - 1)(4\nu^2 -8\nu + 3)(2\nu - 1)(16\nu^4 - 32\nu^3 -\nonumber\\
& &140\nu^2 + 456\nu + 1736)\biggl]\biggl\} ,
\end{eqnarray}
we obtain from (\ref{stephan1}),
\begin{equation}
E^{0}_{k}=\frac{4k}{\pi}\eta^{\gamma-1} \, .
\end{equation} 
\par
On the other hand, using
\begin{eqnarray}
H^{(1)}_{\nu-1}(x)H^{(2)}_{\nu-1}(x)-H^{(1)}_{\nu}(x)H^{(2)}_{\nu}(x) &=& \frac{2}{\pi x}\biggr\{\frac{-8\nu+4}{8x^2}+\nonumber\\
\frac{3}{2}\frac{(-64\nu^3+96\nu^2+16\nu-24)}{64 x^4}\biggl\} \,,
\end{eqnarray}
and inserting in (\ref{stephan2}) we find
\begin{equation}
E^{1}_{k}=\frac{1}{2\pi k}\eta^{\gamma-3}(2\nu-1)^2 \, .
\end{equation}
\par
In the same way it is possible to show that
\begin{equation}
E^{2}_{k}= \frac{2\nu - 1}{2\pi k^3}f(\nu)\eta^{\gamma - 5} \,\, ,
\end{equation}
where
\begin{eqnarray}
 f(\nu) &=&  \frac{1}{12288}\biggr(128\nu^9 - 576\nu^7 - 256\nu^7 + 5968\nu^6 +2127\nu^5
\nonumber\\
&-& 102040\nu^4 + 14432\nu^3 + 554155\nu^2 - 1978\nu - 10794\biggl) .
\end{eqnarray}
\par
From (\ref{expan1},\ref{expan2}) it can be written,
\begin{equation}
E_{k}=\frac{4k}{\pi}\eta^{\gamma-1}+\frac{(2\nu-1)^2}{2\pi k}\eta^{\gamma-3}+\frac{48\nu^4-96\nu^3+24\nu^2+24\nu-9}{32\pi k^3}\eta^{\gamma-5} \, .
\end{equation}
Consequently
\begin{equation}
E^{\,ren}=E_{k}-E^{0}_{k}-E^{1}_{k}-\frac{1}{2}E^{2}_{k}=\frac{h(\nu)}{349152\pi k^3}\eta^{\gamma-5} \,\, ,
\end{equation}
where
\begin{eqnarray}
 h(\nu) &=& - 256\nu^{10} + 1280\nu^9 - 64\nu^8 - 12192\nu^7 - 3657\nu^6 \nonumber\\
 &-& 225352\nu^5 - 57176\nu^4 - 1241334\nu^3\nonumber\\
&+& 5954975\nu^2 + 37044\nu - 14903 .
\end{eqnarray}
\par
Finally, we find
\begin{equation}
\rho^{\,ren}=\frac{h(\nu)\eta^{\gamma-5}}{49152\pi}\int_{0}^{\infty}\frac{dk}{k} \, \, ,
\end{equation}
revealing that a logarithmic divergence remain.

\subsection{Computation for the pressure}

The expression for the pressure is
\begin{equation}\label{igor}
P_{k}=\eta^{\gamma}k^{2}\left\{H_{\nu-1}^{(1)}(k\eta)H_{\nu-1}^{(2)}(k\eta)-\frac{1}{3}H_{\nu}^{(1)}(k\eta)H_{\nu}^{(2)}(k\eta)\right\} \,\,.
\end{equation}
From the definition (\ref{equa1}) it comes out,
\begin{equation}
P_{k}^{n}=\eta^{\gamma}k^2 n\left\{H_{\nu-1}^{(1)}(x)H_{\nu-1}^{(2)}(x)-\frac{1}{3}H_{\nu}^{(1)}(x)H_{\nu}^{(2)}(x)\right\} \,\, ,
\end{equation}
where  $x= nk\eta $.
Hence, 
\begin{eqnarray}
P_{k}^{\,0}&=&\lim_{x\rightarrow\infty}\eta^{\gamma-1}k\,  x\biggr\{H_{\nu-1}^{(1)}(x)H_{\nu-1}^{(2)}(x) -\frac{1}{3}H_{\nu}^{(1)}(x)H_{\nu}^{(2)}(x)\biggl\},\\
P_{k}^{1} &=&\lim_{x\rightarrow\infty}-
\frac{\eta^{\gamma-3}\,x^3}{6k}\biggr\{(2\nu-1)\biggr[3H_{\nu-1}^{(1)}(x)H_{\nu-1}^{(2)}(x)
+ H_{\nu}^{(1)}(x)H_{\nu}^{(2)}(x)\biggl]\nonumber\\
&-& 4x\biggr[H_{\nu}^{(1)}(x)H_{\nu-1}^{(2)}(x)
+ H_{\nu-1}^{(1)}(x)H_{\nu}^{(2)}(x)\biggl]\biggl\} \,\, ,\\
P_{k}^{2}&=&\lim_{x\rightarrow\infty} \frac{\eta^{\gamma-5}x^5}{12k^3}\biggr\{(12\nu^2-3)H_{\nu-1}^{(1)}(x)H_{\nu-1}^{(2)}(x)
\nonumber\\
&+&
(-4\nu^2+8\nu-3)H_{\nu}^{(1)}(x)H_{\nu}^{(2)}(x) \nonumber\\
&+&x(-4\nu-10)\biggr[H_{\nu}^{(1)}(x)H_{\nu-1}^{(2)}(x)
+H_{\nu-1}^{(1)}(x)H_{\nu}^{(2)}(x)\biggl] \nonumber\\
&-&8x^2\biggr[H_{\nu-1}^{(1)}(x)H_{\nu-1}^{(2)}(x)-
H_{\nu}^{(1)}(x)H_{\nu}^{(2)}(x)\biggl]\biggl\}.
\end{eqnarray}

Using (\ref{expan1}) and (\ref{expan2}) we find,
\begin{eqnarray}
H_{\nu-1}^{(1)}(x)H_{\nu-1}^{(2)}(x)-\frac{1}{3}H_{\nu}^{(1)}(x)H_{\nu}^{(2)}(x)=\frac{2}{\pi x}\biggr\{\frac{2}{3}\nonumber\\
+\frac{4\nu^2-12\nu+5}{12 x^2}+\frac{16\nu^4-96\nu^3+104\nu^2+24\nu-27}{64 x^4}\biggl\} \,\, ,
\end{eqnarray}
leading to
\begin{equation}
P_{k}^{\,0}=\frac{4}{3\pi}\eta^{\gamma-1}k \,\, .
\end{equation} 

In a similar way,  we find
\begin{eqnarray}
P_{k}^{1}&=& \frac{4\nu^2 - 12\nu + 5}{6\pi k}\eta^{\gamma - 5}, \\
P_{k}^{2}&=&\frac{g(\nu)}{73728\pi k^3}\eta^{\gamma-5},
\end{eqnarray}
where
\begin{eqnarray}
 g(\nu) &=& 256\nu^{10} - 512\nu^9 - 3456\nu^8 + 12672\nu^7 + 239168\nu^6 - 642912\nu^5\nonumber\\
&-& 278328\nu^4 - 67762\nu^3 - 246453\nu^2 - 150032\nu + 29838 .
\end{eqnarray}
\par
From (\ref{igor}) we have,
\begin{equation}
P_{k}=\frac{4}{3\pi}\eta^{\gamma-1}k + \frac{4\nu^2-12\nu+5}{6\pi k}\eta^{\gamma-3}+\frac{16\nu^4-96\nu^3+104\nu^2+24\nu-27}{32\pi k^3}\eta^{\gamma-5} \,\, .
\end{equation}
Consequently,
\begin{equation}
P^{\,ren}=P_{k}-P_{k}^{0}-P_{k}^{1}-\frac{1}{2}P_{k}^{2}=\frac{l(\nu)}{147456\pi k^3}\eta^{\gamma-5}\,\, ,
\end{equation}
where
\begin{eqnarray}
 l(\nu) &=& - 256\nu^{10} + 512\nu^9 + 3456\nu^8 - 12672\nu^7 - 239168\nu^6 + 642912\nu^5 \nonumber\\
&-& 204600\nu^4 + 235256\nu^3 + 725685\nu^2 + 260624\nu - 154254 .
\end{eqnarray}
\par
The renormalized pressure takes again the form,
\begin{equation}
p^{\,ren}=\frac{l(\nu)}{31474456\pi}\eta^{\gamma-5}\int\frac{dk}{k} \,\, ,
\end{equation} 
with also a logarithmic divergence.

\section{Renormalizing the general inflationary case}

The presence of the logarithmic divergence in the original expression for the
energy and pressure seems to be a general feature of the problem of particle production in cosmology.
See, for example, the reference \cite{birrel}. One important aspect concerning
the logarithmic divergence is that the method of renormalization employed seems inefficient
in the sense that not only it does not eliminate the logarithmic divergence in the
ultraviolet limit, but also it adds a logarithmic divergence in the infrared limit.
Hence, we propose an alternative scheme to cope with this indesirable feature.
\par
The proposed procedure is the following.
First of all, we write the expressions for the energy and pressure as
\begin{eqnarray}
\label{energiabis}
\rho &=& A\eta^{2\nu - 2\beta - 5}\int_0^\infty x^4\biggr\{H_{\nu - 1}^{(1)}(x)\,H_{\nu - 1}^{(2)}(x) + H_\nu^{(1)}(x)\,H_\nu^{(2)}(x)\biggl\}dx, \\
\label{pressaobis}
p &=& A \eta^{2\nu - 2\beta - 5}\int_0^\infty x^4\biggr\{H_{\nu - 1}^{(1)}(x)\,H_{\nu - 1}^{(2)}(x) -  \frac{1}{3}H_\nu^{(1)}(x)\,H_\nu^{(2)}(x)\biggl\}dx,
\end{eqnarray}
where $x = k\eta$. This simple redefinition is well justified if $\eta \neq 0$, that is,
the energy and pressure are computed out of the singularity. But the computation can
be carried out as near of the singularity as we want.
Hence, we have
\begin{eqnarray}
\label{energiabis}
E_k &=& A\eta^{2\nu - 2\beta - 5}x^2\biggr\{H_{\nu - 1}^{(1)}(x)\,H_{\nu - 1}^{(2)}(x) + H_\nu^{(1)}(x)\,H_\nu^{(2)}(x)\biggl\} , \\
\label{pressaobis}
P_k &=& A \eta^{2\nu - 2\beta - 5}x^2\biggr\{H_{\nu - 1}^{(1)}(x)\,H_{\nu - 1}^{(2)}(x) -  \frac{1}{3}H_\nu^{(1)}(x)\,H_\nu^{(2)}(x)\biggl\} .
\end{eqnarray}
The divergencies can then be written as 
\begin{eqnarray}
E_k &=& A\eta^{2\nu - 2\beta - 5}\frac{2}{\pi}x\biggr\{2 + \frac{(2\nu - 1)^2}{4\,x^2} \nonumber\\
&+& 3\frac{(2\nu - 3)(2\nu - 1)^2(2\nu +1)}{64\,x^4} + \cdot\cdot\cdot\biggl\} , \\
P_k &=& A\eta^{2\nu - 2\beta - 5}\frac{2}{\pi}x\biggr\{\frac{2}{3} + \frac{4\nu^2 - 12\nu + 5}{12\,x^2} \nonumber\\
&+& \frac{(2\nu - 3)(2\nu - 1)(2\nu - 9)(2\nu +1)}{64\,x^4} + \cdot\cdot\cdot\biggl\} .
\end{eqnarray}
 We subtract the (divergent) expressions for
energy and pressure as
\begin{eqnarray}
 E_k^{ren} &=& E_k - E_k^0 - E_k^2 - E_k^{log} , \\
P_k^{ren} &=& P_k - P_k^0 - E_k^2 - P_k^{log} ,
\end{eqnarray}
with the definitions 
\begin{eqnarray}
 E_k^{log} &=& A\eta^{2\nu - 2\beta - 5}\frac{6}{\pi}\frac{(2\nu - 3)(2\nu - 1)^2(2\nu +1)}{64\,x^3}(1 - e^{-\sigma x}) , \\
P_k^{log} &=& A\eta^{2\nu - 2\beta - 5}\frac{2}{\pi}\frac{(2\nu - 3)(2\nu - 1)(2\nu - 9)(2\nu +1)}{64\,x^3}(1 - e^{-\sigma x}) ,
\end{eqnarray}
where an extra term has been added in order to assure that the divergence is eliminated
in the ultraviolet limit without creating a new divergence in the infrared limit.
The parameter $\sigma$ defines the efficiency of this mechanism, and the final result must
be independent of it in order to guarantee the consistency of the procedure proposed.
\par
The energy and the pressure are now given by
\begin{eqnarray}
\rho^{ren} &=& \int_0^\infty x^2(E_k - E_k^0 - E_k^1)dx - \int_{1/\sigma}^\infty x^2 E_k^{log}dx , \\
p^{ren} &=& \int_0^\infty x^2(P_k - P_k^0 - P_k^1)dx - \int_{1/\sigma}^\infty x^2 P_k^{log}dx  .
\end{eqnarray}
\par
Essentially we must compute an integral of the type
\begin{equation}
 I = \int_{1/\sigma}^\tau \frac{1 - e^{-\sigma x}}{x}dx ,
\end{equation}
showing that the logarithmic divergence can be eliminate in the ultraviolet limit,
when $\tau \rightarrow \infty$, remaining
a final result independent of $\sigma$. In fact, this expression can be defined
as
\begin{equation}
 I = \int_{1}^\tau \frac{1 - e^{-y}}{y}dy ,
\end{equation}
where $y = \sigma x$. The result is
\begin{equation}
 I = \ln \tau - \int_1^\tau\frac{e^{-y}}{y}dy  .
\end{equation}
The first term eliminates the ultraviolet logarithmic divergence in the limit
$\tau \rightarrow \infty$ while the second one remains finite in the same limit.
Moreover, the result is independent of $\sigma$. This procedure can be easily generalized
if the cut-off is introduced as
\begin{equation}
 I = \int^\tau_{1/\sigma^{1/n}} (1 - e^{-\sigma x^n})\frac{dx}{x} ,
\end{equation}
where $n$ is any positive number.
\par
As a result, we remain with the expressions for the energy and pressure
\begin{eqnarray}
 \rho^{ren} &=& \bar A_1\eta^{2\nu - 2\beta - 5} I_1 = \bar A_1\eta^\frac{-12(1 + \alpha)}{1 + 3\alpha} I_1 , \\
p^{ren} &=& \bar A_2\eta^{2\nu - 2\beta - 5} I_2 = \bar A_2\eta^\frac{-12(1 + \alpha)}{1 + 3\alpha} I_2 ,
\end{eqnarray}
where $\bar A_1$ and $\bar A_2$ are constants and $I_1$ and $I_2$ are the (finite)
integrals resulting from the renormalization procedure sketched before.
\par
Notice that, since the Ricci scalar can be written
as
\begin{equation}
 R = 6 \frac{a''}{a^3} \propto \eta^{- 6\frac{1 + \alpha}{1 + 3\alpha}} ,
\end{equation}
the pressure and energy density are proportional to the square of the Ricci curvature.
Hence, when the null energy condition is violated ($\alpha < - 1$), the energy of the
created particles diverges as the singularity is approached, while in the case
the strong energy condition is violated but the null energy condition is satisfied,
the energy of the created particle is very high in the beginning decreasing as the
the universe expands.
\par
Hence, using the renormalization procedure proposed here, the energy of created particle
may be enough to avoid the big rip, in opposition to what has been stated in \cite{brasil1}.
In fact the ratio of the renormalized energy of the created particle to the energy density
of the dark energy fluid $\rho_x$ is
\begin{equation}
 \frac{\rho^{ren}}{\rho_x} \propto \eta^{-6\frac{1 + \alpha}{1 + 3\alpha}},
\end{equation}
going to zero when $ - 1/3 > \alpha > - 1$ and to infinity  when $\alpha < - 1$.
The case of the de Sitter space-time $\alpha = - 1$, is just the limiting one, the
energy of the created particle neither decreasing nor incresing, remaining always equal to
zero.

\section{Conclusions}

In this work we have re-analyzed the problem of particle creation in a universe dominated
by dark energy. Special attention has been given to the case where dark energy is
represented by a phantom fluid. This problem has already been addressed in reference
\cite{brasil1}. There the conclusion was that the energy associated to the created particles,
as the universe approaches the big rip, goes to zero in spite of the fact that the
number of created particles goes to infinity.
\par
The conclusion presented in reference \cite{brasil1} has been obtained by introducing a cut-off
in the energy integral at Planck's scale. This seems to be natural because it can be
expected that the usual Klein-Gordon equation must be modified beyond the Planck's scale,
and some proposals in this sense suggest the introduction of an exponential decreasing
term with the wavenumber $k$, suppressing the contribution due to the transplanckian
scales \cite{lemoine}.
\par
Here, instead of introducing an upper cut-off, we have chosen to renormalize the energy integral using the $n$ wave method described
in reference \cite{zel}. First we have shown that the employement of this method leads to the
expected conclusions concerning the radiative and de Sitter universes: the energy
expression is divergent, but after renormalization it becomes zero.
\par
Applying the method to the general dark energy model, a difficulty appears due to
a logarithmic divergence in the ultraviolet limit. We have proposed a method to
cope with this difficulty, introducing a new cut-off. The final result is independent
of the cut-off. Moreover, the final expressions for energy and pressure are
covariantly conserved.
\par
The result for the energy density differs from that found in reference \cite{brasil1},
since now the energy density of the created particles diverges as the the big rip
is approached, and becomes the dominant component. Hence, following this result, quantum
effects can be effective to avoid the big rip. In fact, the renormalized energy
comes out to be proportional to the square of the Ricci scalar. Another consequence of the
computation made in the present work is that in the
non phantom dark energy models the renormalized energy goes asymptotically to zero.
The de Sitter case is the separatrix of these two different behaviour.
\par
An important point concerning this result is its unicity. A comparison must be made with
other renormalization methods, for example those described in
\cite{pavlov1} and references therein. We hope to address this problem in a future work. However,
the absence of any final dependence on the cut-off employed to renormalize the energy,
together with the fact that the renormalized energy and pressure obeys a covariant
conservation law, indicates that this result is quite consistent.
\newline
\vspace{0.5cm} 
\newline
{\bf Acknowledgement:} We thank CNPq (Brazil), FAPES (Brazil) and the brazilian-french scientific cooperation CAPES/COFECUB for partial
financial support. F.G.A. and J.C.F. thank $Gr\epsilon CO$, IAP, France, for kind
hospitality during part of elaboration of this work. We thank Olivier Piguet for the
critical reading of the manuscript.

\end{document}